# Does A Paper Being Featured on The Cover of A Journal Guarantee More Attention and Greater Impact?

Xianwen Wang*, Chen Liu and Wenli Mao

WISE Lab, School of Public Administration & Law, Dalian University of Technology, Dalian 116085, China.

* Corresponding author.
Email address: xianwenwang@dlut.edu.cn; xwang.dlut@gmail.com

Website: xianwenwang.com

**Abstract**
Paper featured on the cover of a journal has more visibility in an issue compared with other ordinary articles for both printed and electronic journal. Does this kind of visibility guarantee more attention and greater impact of its associated content than the non-cover papers? In this research, usage and citation data of 60 issues of *PLOS Biology* from 2006 to 2010 are analyzed to compare the attention and scholarly impact between cover and non-cover paper. Our empirical study confirms that, in most cases, the group difference between cover and non-cover paper is not significant for attention or impact. Cover paper is not the best one, nor at the upper level in one issue considering the attention or the citation impact. Having a paper featured on the cover of a journal may be a source of pride to researchers, many institutions and researchers would even release news about it. However, a paper being featured on the cover of a journal doesn't guarantee more attention and greater impact.
**Keywords:** *Paper featured on cover; Cover image; Cover paper; Article usage; Article view; Citation*

# Introduction

In the era of print, cover image is the first thing to catch people's sight when someone picks up a journal. With the fast development of digital libraries and gradual disappearance of print, readers have much less possibilities to touch the physical copy of a journal issue. However, cover of a journal is still more visible than any other pages inside. Many publishers still keep the cover image and cover story. Journal cover with well-designed image is always placed at an eye-catching location on the journal website.

There are different opinions on the selection to feature a paper on the cover of journal (cover paper). For some journals, they choose the very best paper of an issue on the cover, "a paper that in 20 years' time might win a Nobel Prize," according to the opinion of Stang, the chief editor of *Journal of the American Chemical Society* (Ritter, 2006). However, for many other journals, the selection standard could be completely different. For example, the selection to feature a particular article on the cover of *Nature Chemistry* does not imply that it is better than the other papers in the issue. The reason why an article is chose as cover paper just because of its eye-catching (Small, 2004), or striking image (Nature Chemistry, 2010), which also applies to *PLOS Biology*, our research object in this study (http://www.plosbiology.org/static/checklist). For *Emerging Infectious Diseases* journal, images for the cover are selected for "artistic quality, technical reproducibility, stylistic continuity, communication effectiveness, and audience appeal", except the quality or importance of the study (Emerging Infectious Diseases, 2013).

Having a paper featured on the cover of a journal is also a source of pride to researchers, it may attract admiration from peers and offer increased exposure to researchers and their work (Nature Chemistry, 2010; Ritter, 2006). Many institutions and researchers would even release news about it. Some funding agencies, such as the National Science Foundation and the ACS Petroleum Research Fund, typically react favorably if the material has been featured on international journal covers (Ritter, 2006).

Our research question is that, given that the selection of cover story has nothing to do with the quality of the paper, does being a cover paper guarantee more attention and greater impact of its associated content than the non-cover papers? In this study, article views data is used to represent attention, when citation stands for academic impact.

## Data source

In this study, *PLOS Biology* is selected as our research object. *PLOS Biology* is an open-access, peer-reviewed general biology journal published by PLOS, issues are published monthly. Figure 1 shows some cover images of *PLOS Biology*.

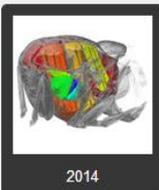

Figure 1 Screenshot of cover image of *PLOS Biology*

In March 2009, PLOS started an article-level-metrics program to provide metrics data for every article published as early as 2003 across all PLOS journals. The metrics indicators include four categories, which are viewed, cited, saved and discussed. Compared with the traditional bibliometrics indicators, e.g., citation, the PLOS article-level-metrics (ALMs) provides "a wide range of metrics about the uptake of an individual journal article by the scientific community after publication" (Fenner, 2013). In recent years, more and more academic publishers and scholarly journals begin to provide ALMS to public, e.g., Nature journals (in December, 2012), *PNAS* (in January, 2014) (Wang et al. 2014).

On January 25 and 26, 2014, article-level-metrics data for all research articles (as opposed to reviews, editorial materials, letters, or news features) published from 2006 to 2010 are harvested directly from the website of *PLOS Biology*. For each article, the total views (regarded as attention) and Web of Science citations data (regarded as scholarly impact) are parsed and extracted from the metrics webpage. Finally, data of 1025 articles (60 cover papers and 965 non-cover papers) is collected, processed and then imported into our designed SQL Server database to make analysis.

In the dataset, the metrics (article views and citations) window is from the date of publication until date of data harvesting. Articles published earlier may profit from a larger metrics window. To avoid this kind of bias, besides the holistic analysis, we also make analysis of the data annually, which means that in the annual subdataset, data are partitioned by the article publish year, only those articles published in the same year are compared together. For different annual subdatasets, the metrics windows are different, e.g., for the subdataset of 2006, the window is 7 years; and for the subdataset of 2010, the window is 3 years.

## Results

*Scatter plot*

Figure 2 is the scatter plot that displays the views (x axis, attention) and citation (y axis, impact) for the whole dataset. The red dots represent the cover paper, when the gray dots are non-cover papers. Most dots are distributed in a small area close to the origin of the axes, the red and grey dots are mixed together, it is hard to tell that cover papers has much advantage in views and citations. There are also some outliers distant from the majority of observations, which are fenced by an irregular dotted box in Figure 2.

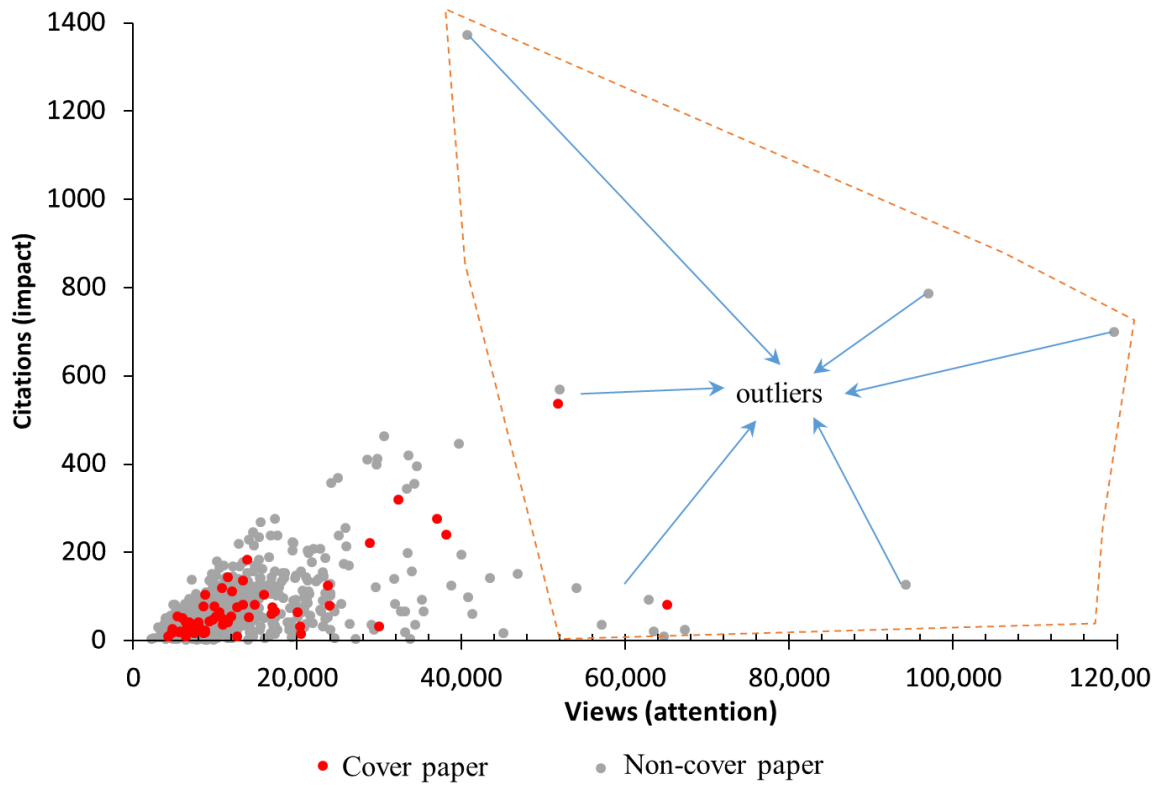

Figure 2 Scatter plot of views and citation of cover and non-cover paper

*Position of cover and non-cover paper in one issue*

To better compare the position of cover and non-cover paper in the same issue from both the prospects of attention and impact, all data is visualized as Figure 3 shows. Papers published in the same issue are arranged in one column, 60 columns represent 60 issues published from 2006 to 2010. Articles in each year are placed in each corresponding panel of Figure 3 (a) to (e) with different colors. The scale range is different for the 5 panels. Cover papers in all years are highlighted in red solid circle, when empty circles with different colors denote non-cover papers in the corresponding year. In all panels, the y axis denotes the views, when the size of circles measures the citations, which means that big circles with high position have more views and citations than the lower smaller ones in the same column.

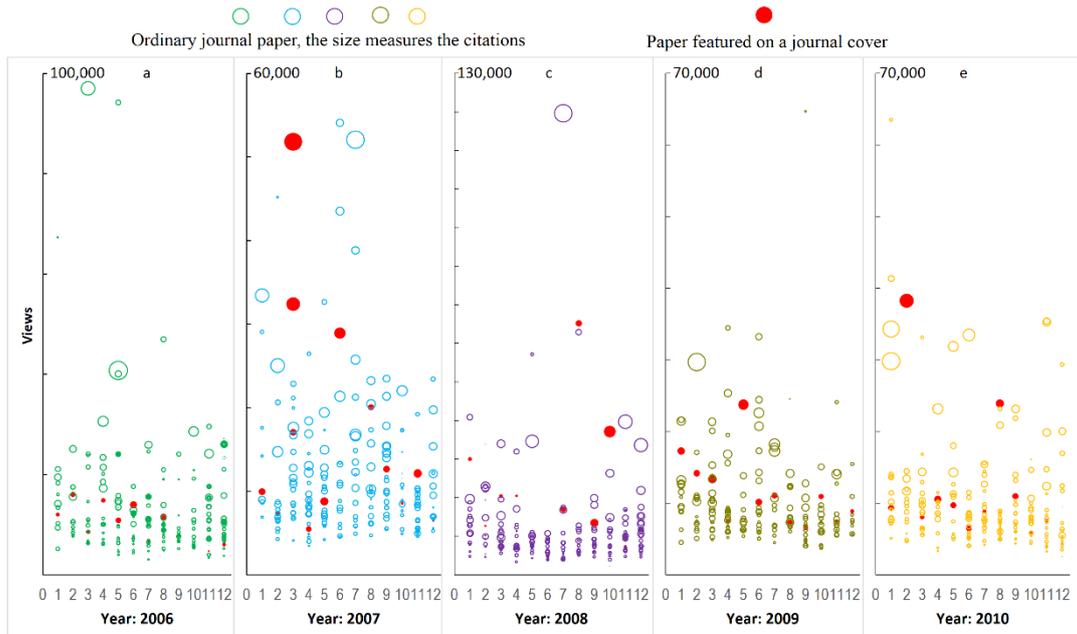

Figure 3 Annual comparisons of views and citations between cover and non-cover papers

Two issues (July 2007 and May 2008) don't have cover papers, and three papers are featured on the single issue in March 2007. For the other 58 issues (columns) which have cover papers, as Figure 3 shows, only five columns have red circles ranked top one (the highest position in the column), and 21 columns have red circles ranked top three. Moreover, four columns even have red circles at the lowest position (last three). For the citations, only seven issues have cover papers with the most citations, 14 issues have cover papers with top three most citations, and even five issues have cover papers with the last fewest citations (last three). As Table 1 shows, only 36.21% of cover papers have obvious advantage of article views, for citations, the ratio is even lower, only 24.14% have obvious advantage.

Table 1 The number of issues with cover paper at the specific position

|  | views | Percentage | citations | Percentage |
|---|---|---|---|---|
| Top one | 5 | 8.62% | 7 | 12.07% |
| Top three | 21 | 36.21% | 14 | 24.14% |
| Last three | 4 | 6.90% | 5 | 8.62% |

*One Way ANOVA Analysis*

A one-way ANOVA is used to evaluate the effect of cover featuring on the views and citations. The alpha level is set at 0.01 instead of 0.05 to avoid possible Type I error because of the repeated tests (over year and adjusted). ANOVA is calculated on the difference of views and citations between cover and non-cover paper. For citations, $F(1, 1023) = 1.614$, $p = .204$; when for views, $F(1, 1023) = 6.095$, $p = .014$, as Table 2 shows. Neither the analysis for citations nor views is significant, which means that there is not significant difference between cover and non-cover papers. More specifically, annual ANOVA analysis is conducted for the subdataset of each year. In most years, the analysis for both views and citations between cover and non-cover paper is not significant. However, the analysis for citations in 2007, and the analysis for views in 2008 are significant. We think that some outliers may have bias on the overall result. As shown in Figure 1, the outliers distant from the majority of observations are fenced.

Therefore, the outliers (two cover papers and eleven non-cover papers) are excluded. Annual ANOVA analysis is conducted again for the rest data of the subdataset from 2006 to 2010. Moreover, the whole dataset from 2006 to 2010 without the excluded data mentioned above is also analyzed with ANOVA analysis, as shown in Table 2, the rows in boldface denote the result after adjustment.

For the annual ANOVA analysis of the adjusted data, all the results except the views in 2008 confirm no significant group differences in neither views nor citations, e.g., the analysis for citations in 2007 is significant, $F(1, 218) = 7.949$, $p = .005$; after the adjustment, the analysis is not significant, $F(1, 215) = 2.595$, $p = .109$. However, for views in 2008, the analysis is always significant whether no matter before or after the adjustment. For the whole dataset, after the first adjustment, ANOVA still confirms no significant group differences in views, $F(1, 1011) = 6.213$, $p = .013$; and in citations, $F(1, 1011) = .604$, $p = .437$ (see Table 2).

Table 2 ANOVA analysis

|  | Mean (SD) | | F | Significance |
|---|---|---|---|---|
|  | Cover paper | Non-cover paper | | |
| all_views | 14378.37 (11430.899) | 11249.54 (9395.644) | 6.095 | p = .014 |
| **all_views adjusted** | **12856.21 (7934.183)** | **10600.18 (6611.137)** | **6.213** | **p = .013** |
| all_citations | 74.58 (88.544) | 60.99 (79.865) | 1.614 | p = .204 |
| **all_citations adjusted** | **66.59 (65.931)** | **59.05 (71.991)** | **.604** | **p = .437** |
| 2006_ views | 9918.58 (3828.352) | 13290.46 (12266.014) | 0.897 | p = .345 |
| **2006_ views adjusted** | **9918.58 (3828.352)** | **12011.80 (7386.990)** | **.942** | **p = .333** |
| 2006_ citations | 75.33 (51.286) | 98.63 (131.176) | 0.373 | p = .542 |
| **2006_ citations adjusted** | **75.33 (51.286)** | **94.90 (121.193)** | **.308** | **p = .580** |
| 2007_ views | 17052.85 (13443.393) | 11649.97 (7904.959) | 5.175 | p = .024 |
| **2007_ views adjusted** | **14154.67 (8833.739)** | **11245.8341 (6789.985)** | **2.009** | **p = .158** |
| 2007_ citations | 128.69 (148.764) | 69.08 (67.070) | 7.949 | p = .005[***] |
| **2007_ citations adjusted** | **94.92 (89.242)** | **66.42 (57.526)** | **2.595** | **p = .109** |
| 2008_ views | 21792.27 (17255.556) | 11387.37 (11225.557) | 8.405 | p = .004[***] |
| **2008_ views adjusted** | **17450.40 (10021.298)** | **10338.62 (6509.176)** | **10.710** | **p = .001[***]** |
| 2008_ citations | 64.45 (78.31) | 62.05 (75.656) | 0.01 | p = .919 |
| **2008_ citations adjusted** | **63.00 (82.392)** | **58.81 (60.981)** | **.043** | **p = .835** |
| 2009_ views | 11542.75 (5028.989) | 10054.77 (6757.889) | 0.56 | p = .455 |
| **2009_ views adjusted** | **11542.75 (5028.989)** | **9754.45 (5415.260)** | **1.237** | **p = .267** |
| 2009_ citations | 46.25 (30.666) | 44.36 (38.631) | 0.028 | p = .868 |
| **2009_ citations adjusted** | **46.25 (30.665)** | **44.55 (38.643)** | **.022** | **p = .882** |
| 2010_ views | 11980.33 (9638.092) | 10027.81 (7600.706) | 0.725 | p = .396 |
| **2010_ views adjusted** | **11980.33 (9638.092)** | **9761.57 (6608.291)** | **1.206** | **p = .273** |
| 2010_ citations | 52.83 (61.707) | 34.33 (43.816) | 1.922 | p = .167 |
| **2010_ citations adjusted** | **52.83 (61.707)** | **34.41 (43.912)** | **1.897** | **p = .170** |

[***] Significant at p < 0.01

# Conclusion

With the movement from print to electronic publishing, cover image is still more visible than any other pages inside. Although journal cover features bring pride to researchers and may increase the visibility and exposure to researchers and their work, our empirical study find that a cover is just a cover, it would not increase the attention nor the scholarly impact of the study. In most cases, cover paper is not the best one, nor at the upper level in an issue considering the attention or the citation impact. However, as a combination of art and science, cover art image has become a kind of culture of online science communication (Liu, 2013), it means a lot to the readers, the publishers and science community. With the mission to communicate the complexity and beauty of science, even the paper journals may draw to a close someday, the concept of cover may not (Nature Chemistry, 2010).

# Acknowledgements

The work was supported by the project of ''National Natural Science Foundation of China'' (61301227), and the project of "Growth Plan of Distinguished Young Scholar in Liaoning Province"(WJQ2014009).